# High-frequency behavior of FeN thin films fabricated by reactive sputtering

Tae-Jong Hwang, Joonsik Lee, Ki Hyeon Kim, and Dong Ho Kim

Department of Physics, Yeungnam University, Gyeongsan, 38541, Korea

**Abstract**

We investigated high-frequency behavior of FeN thin films prepared by reactive sputtering through ferromagnetic resonance (FMR) and its relationship with the static magnetic properties. The FMR was observed in the frequency range from 2 to 18 GHz in the FeN films fabricated at proper nitrogen flow rate (NFR). In those FeN thin films, a decrease of the saturation magnetization and the corresponding decrease of the FMR frequency were observed as NFR was increased during the deposition. The external field dependences of the FMR frequencies were well fit to the Kittel formula and the Landé g-factors determined from the fit were found to be very close to the free electron value. The high-field damping parameters were almost insensitive to the growth condition of NFR. However, the low-field damping parameters exhibited high sensitivity to NFR very similar to the dependence of the hard-axis coercivity on NFR, suggesting that extrinsic material properties such as impurities and defect structures could be important in deciding the low-field damping behavior.

Corresponding author: Dong Ho Kim

Email: dhkim@ynu.ac.kr

## Introduction

When a magnetic material with magnetization $M$ is placed in an external magnetic field $H$ and $M$ and $H$ are initially not parallel, a torque is exerted on the magnetization causing a precession around the external field direction. The precession frequency is given by so called the Lamor frequency, $\gamma H/2\pi$, where $\gamma$ is the gyromagnetic ratio defined as $\gamma = \frac{g\mu_B}{\hbar}$. Here g is the Landé g-factor and $\mu_B$ is the Bohr magneton. The precession motion will continue forever when there is no damping force. However, in the real materials, the magnetization will eventually align to the minimum energy direction as the energy is transferred from the precession motion to the environment through spin-spin or spin-lattice interaction. In the small damping limit, the motion of magnetization subject to an effective magnetic field $H_{eff}$ is described by the Landau-Lifschitz-Gilbert (LLG) equation [1]

$$\frac{d\mathbf{M}}{dt} = -\gamma(\mathbf{M} \times \mathbf{H}_{eff}) + \frac{\alpha}{M_s}\mathbf{M} \times \frac{d\mathbf{M}}{dt}, \qquad (1)$$

where $4\pi M_S$ is the saturation magnetization and $\alpha$ is the Gilbert damping parameter. Due to the anisotropy contributions, $H_{eff}$ is sum of the anisotropy field $H_K$ and applied external magnetic field $H_{ext}$. Solving equation (1) with $H_{ext}$ applied along the magnetization easy axis and a small uniform excitation field perpendicular to $H_{ext}$ gives complex magnetic susceptibility of the material. Imaginary part of susceptibility, corresponding to energy absorption by the magnetic material, becomes maximal at a frequency defined as the resonance frequency $f_{res}$. This phenomenon is named ferromagnetic resonance (FMR) and $f_{res}$ is written by the Kittel formula [2]

$$f_{res} = \frac{\gamma}{2\pi}\sqrt{(H_{ext} + H_k)(H_{ext} + H_k + 4\pi M_s)}\,. \qquad\qquad (2).$$

With known $4\pi M_s$, the field dependence of $f_{res}$ can be fit to equation (2) to obtain $H_K$ and $\gamma$. Using the linewidth $\Delta f$ of the resonance peak, the Gilbert damping parameter $\alpha$ can be estimated by the following relation

$$\alpha = \frac{2\pi\Delta f}{\gamma[2(H_{ext} + H_k) + 4\pi M_s]}\,. \qquad\qquad (3)$$

The iron-nitrogen system has been studied for many decades because of the excellent mechanical and magnetic properties. Especially, effect of nitrogen inclusion on the structural and magnetic properties of FeN films have been one of the mostly investigated subjects. [3-10] On the other hand, studies on high-frequency behaviors of FeN film have been seldom performed. [9,10] In this study, we report the results of intensive study of high-frequency characteristics of FeN thin films in which nitrogen contents were systematically varied through the deposition process. We obtained high-frequency parameters such as FMR frequencies, damping parameters, and g-factors, and discuss their relationship with static magnetic properties, electrical resistivity and crystal structures.

**Experimental**

The FeN thin films were fabricated on single-crystal Si (100) substrates by reactive RF magnetron sputtering. An external field of 400 Oe was applied to the in-plane direction in order to induce in-plane uniaxial anisotropy. The base pressure of the sputtering chamber was below $3 \times 10^{-7}$ Torr. The sputtering gas was mixture of Ar and $N_2$ and their mixing ratio was controlled by the flow rate of each gas. The Ar flow rate was fixed at 10.0 sccm while $N_2$ flow rate (NFR) was varied from 1.1 to 2.1 sccm with 0.1 sccm step to adjust the nitrogen content in FeN thin films. For any $N_2$/Ar flow-rate ratio, the total pressure of the gas mixture was maintained at 3 mTorr. The sputtering power was 100 W and the substrate holders were cooled with coolant at 5 °C. The deposition time was adjusted to fabricate 100 nm thick films and the actual thicknesses of the films measured by using a surface profiler were 100~105 nm. The resistivity was measured by using the four-probe method.

The crystal structure was analyzed by X-ray diffraction (XRD) patterns. Static magnetic properties such as the *M-H* loops were investigated by vibrating sample magnetometer (VSM). High frequency behavior, FMR frequency and frequency linewidth, were obtained by using vector network analyzer (VNA) and grounded coplanar waveguide (CPW) in the frequency range from 100 MHz to 26.5 GHz. An external magnetic field up to 2 kOe has been applied parallel to the magnetization easy axis and the RF field along the hard axis. [11]

**Results and discussion**

Fig. 1 shows the XRD patterns of $\theta$-$2\theta$ scan of representative FeN thin films deposited at different nitrogen gas flow rate. The pure Fe thin film prepared by the same deposition condition showed polycrystalline nature with a rather broad $\alpha$-Fe (110) peak at $2\theta = 44.36°$ in the XRD pattern. With increase of NFR to 1.2 sccm, the (110) peak became much broader and the peak position was slightly shifted to a lower angular position of $2\theta = 44.12°$. For films prepared at NFR of 1.4 sccm, the (110) peak is replaced by just a trace of the peak. With further addition of nitrogen of NFR higher than 2.0 sccm, a new broad peak appeared at $2\theta =$ 43.15°, which is assumed to be a (101) peak of ε-Fe$_3$N phase. [5] Atomic percentages of Fe and N elements were obtained by using X-ray energy dispersion spectroscopy. The nitrogen atom percentage was ~29% for FeN thin films prepared at NFR of 1.1 sccm, and it gradually increased with increasing NFR, becoming ~35% for NFR of 1.4 sccm, and finally reached ~46% at NFR of 2.0 sccm. It means that addition of proper amount of nitrogen induces an amorphous phase, in such, nitrogen atoms are randomly distributed on interstitial sites. But further incorporation of nitrogen yields a partial formation of unwanted phase such as ε-Fe$_3$N.

The electrical resistivity of FeN thin has also been affected by the nitrogen addition. The resistivity of pure Fe film was 10 μΩ·cm, similar to that of the single crystalline Fe film. [4] It increased monotonically reaching 70 μΩ·cm at NFR of 1.3 sccm then saturated around 90 μΩ·cm at higher NFR. The resistivities of the current FeN films are smaller than the reported values of ~200 μ Ω·cm for films prepared with Ar and N$_2$ gas flow rates of 12 and 3 sccm, respectively. [10]

The role of nitrogen addition on the magnetic properties of FeN thin film can be clearly observed in Fig. 2 which shows how *M–H* loops develop with increasing NFR. The solid lines are the loops measured with fields along the easy axis of magnetization and the dotted lines along the hard axis. The pure Fe film exhibited almost isotropic in-plane magnetization and high saturation magnetization $4\pi M_S$ of 23 kG. This value is comparable to or slightly greater than the reported values in the literatures, [5-7,10] implying that our growth condition for Fe thin film is well optimized. Increase of NFR generated monotonic reduction of the saturation magnetization throughout the entire NFR range, consistent with earlier studies. [5,7,8] It has been reported that further increase of nitrogen gas fraction during the growth process eventually incurs a complete disappearance of ferromagnetism in FeN films. [7,8]

In the case of coercivities, non-monotonic dependence on NFR was observed. The coercivities first decreased from 18 Oe of the pure Fe film to reach the minimum value at the film prepared at NFR of 1.4 sccm. The corresponding coercivities for easy and hard axis, $H_{ce}$ and $H_{ch}$, were 1.5 Oe and 1.9 Oe, respectively, as can be seen in Fig. 2(c). With further increase of NFR, $H_{ce}$ started to increase, reaching ~10 Oe for FeN thin films grown at NFR of 1.7 sccm. Beyond NFR of 1.8 sccm, totally different magnetic behavior has been observed. A typical $M$–$H$ loop in this range is shown in Fig. 2(d), where anisotropy exists, but of different shape, and the coercivities are much larger than those of pure Fe thin films. This happens probably due to the appearance of $\varepsilon$-Fe$_3$N phase.

In the films with very small grains of randomly oriented, for example in amorphous phase, the effective anisotropy energy is reduced due to the fact that more grains are involved in the magnetic coupling, which in turn results in the partial cancellation of the crystalline anisotropy. As a result, coercivities would be significantly reduced [12] and samples will exhibit soft magnetic properties. The dependences of $4\pi M_S$, $H_{ce}$ and $H_{ch}$ on NFR of soft and uniaxial in-plane anisotropic FeN thin films are summarized in Fig. 3.

As described earlier, the high-frequency dynamics of magnetization can be well described by the phenomenological LLG formula in equation (1) with a characteristic precession and the Gilbert damping associated with the precession. When the frequency of RF field is equal to $f_{res}$, resonance absorption takes place. In this work, the FMR parameters were obtained from standard microwave $S$-parameter measurements by using VNA and CPW as a function of frequency at various static magnetic fields.

Only the FeN films produced at NFR range from 1.3 to 1.7 sccm, also having uniaxial in-plane anisotropy, showed meaningful FMR signals. The data were analyzed assuming that the dominant CPW signals were in the quasi-transverse electromagnetic mode. Fig. 4(a) shows the amplitude spectra of $S_{21}$ of the FeN thin film prepared at NFR of 1.5 sccm. As the external field was increased from 0 to 2,000 Oe, the resonance curve moved to higher frequencies consistent with the Kittel formula in equation (2); as $H_{ext}$ increases, so does $f_{res}$. The other specimens fabricated at different NER showed a very similar trend in magnetic fields. Representative spectra at 500 and 750 Oe for three films prepared at NFR of 1.3, 1.5, and 1.7 sccm, respectively, are shown in Fig. 4(b). It shows that the spectra shift to lower frequency with increasing NFR. The FMR frequency $f_{res}$ and the frequency linewidth $\Delta f$ at the half maximum of the resonance peak were determined from the fit of $S_{21}$ spectrum to the Lorentzian curve. Before the fit, the background signal from the Si substrate was subtracted. The solid lines in Figs. 4(a) and (b) are the Lorentzian fit.

The $f_{res}$ as a function of $H_{ext}$ for three FeN thin films, for clarity, are plotted in Fig. 5(a). The decrease of $f_{res}$ with increasing NFR is due to the decrease of $M_S$ with increasing NFR (see Fig. 3 and equation (2)). The solid lines in Fig. 5(a) are fit to the Kittel formula. All the $f_{res}$ data showed good fit to equation (2). Since $4\pi M_S$ had been determined from the VSM measurement, $\gamma$ or g-factor and $H_K$ were used for the two fitting parameters. The NFR dependences of the fitting parameters are shown in Fig. 5(b). The g-factors were within 1.98 and 2.06, very close to the free electron value, and $H_K$ were found to be in the range from 30 to 38 Oe. The obtained $H_K$ values from the fit were in a good agreement with those measured from the $M$–$H$ loops, for instance, $H_K$ of 35 Oe from the fit for NFR of 1.4 sccm is comparable to the experimental result in Fig. 2(c). No systematic dependence of either g-factor or $H_K$ on NFR was observed.

The Gilbert damping parameters estimated from $\Delta f$ and equation (3) for five FeN thin films are shown in Fig. 6(a) as a function of external magnetic field. Logarithmic scale was used for field axis to emphasize the low-field behavior. Note that the data points plotted at 0.5 Oe correspond to those measured at a zero external field. All the $\alpha$ decreased with increasing field. The high-field $\alpha$ values for all the samples are almost the same in the range between 0.0052 and 0.0058. These values of ~0.005 are comparable to those observed in as-grown FeCoB films. [13] However, the low-field $\alpha$ values were vastly different from each other, shown as a function of NFR in Fig. 6(b). Before fully saturated, thin films used to have non-uniform magnetization or anisotropy which varies not only from position to position because the sample is not in a uniform magnetic state but from sample to sample since films are fabricated at different deposition conditions. The former is the origin for the enhancement of damping at low fields and the latter is responsible for the sample to sample variation of low-field damping. Note that NFR dependence of low-field $\alpha$ in Fig. 6(b) resembles to those of coercivities in Fig. 3, that is, the higher coercivities the higher low-field $\alpha$ values. More resemblance to $H_{ch}$ is observed presumably because the RF electric field is applied along the hard axis such that energy loss at RF frequency would be affected more by $H_{ch}$. The intrinsic origin of the Gilbert damping has been intensively discussed in terms of spin-orbit coupling. [14,15] However, our finding of g-factor very close to 2.00 indicates that damping due to the spin-orbit coupling is rather weak. Furthermore, because of the fact that the spin-orbit coupling effect would be almost field independent for applied fields < 2 kOe, the spin-orbit effect alone may not explain the enhanced low-field damping. Instead, extrinsic material properties such as impurities and defect structures that strongly depend on deposition conditions could play more important roles in deciding the low-field damping behavior.

**Summary**

We investigated the relationship between high frequency behavior and the static magnetic properties of FeN thin films prepared at various nitrogen flow rates. The saturation magnetization showed a monotonic decrease with increasing NFR, so does the resonance frequency. On the other hand, the damping parameters at low field exhibited strong dependence on the nitrogen flow rate while high-field damping parameters were unaffected. We observed a correlation between the hard-axis coercivities and low-field damping parameters.

**Acknowledgments**

This work was supported by the National Research Foundation of Korea (NRF) grant funded by the Korea government (MSIP) No. 2014R1A2A1A11051500.

**Figure Captions**

Fig. 1. XRD patterns of $\theta$-$2\theta$ scan of representative FeN thin films deposited at various nitrogen gas flow rates.

Fig. 2. Magnetization hysteresis loops of FeN thin films fabricated at the nitrogen flow rates of (a) 0 sccm, (b) 1.2 sccm, (c) 1.4 sccm, and (d) 2.0 sccm. The solid lines are easy-axis loops and the dotted lines are hard-axis loops. Two films of (b) and (c) showed soft magnetic properties as well as uniaxial in-plane anisotropy.

Fig. 3. The dependences of $4\pi M$s, $H_{ce}$ and $H_{ch}$ on the nitrogen flow rate of soft and uniaxial in-plane anisotropic FeN thin films.

Fig. 4. (a) The amplitude spectra of $S_{21}$ measured at various external fields from 0 to 2,000 Oe for the FeN thin film fabricated at the nitrogen flow rate of 1.5 sccm. (b) Representative spectra at 500 and 750 Oe for three films prepared at NFR of 1.3, 1.5, and 1.7 sccm, respectively. It shows that the spectra shift to lower frequency with increasing NFR. The solid lines are fit to the Lorentzian curve.

Fig. 5. (a) External field dependence of the ferromagnetic resonance frequencies of FeN thin films fabricated at 1.3, 1.5, and 1.7 sccm. The solid lines are fit to the Kittel formula. (b) Obtained g-factors and anisotropy fields from the fit.

Fig. 6. (a) Damping parameters of five FeN thin films as a function of external field. (b)

Dependence of the damping parameters on the nitrogen flow rate at external fields of 0, 1,000, and 1,500 Oe.

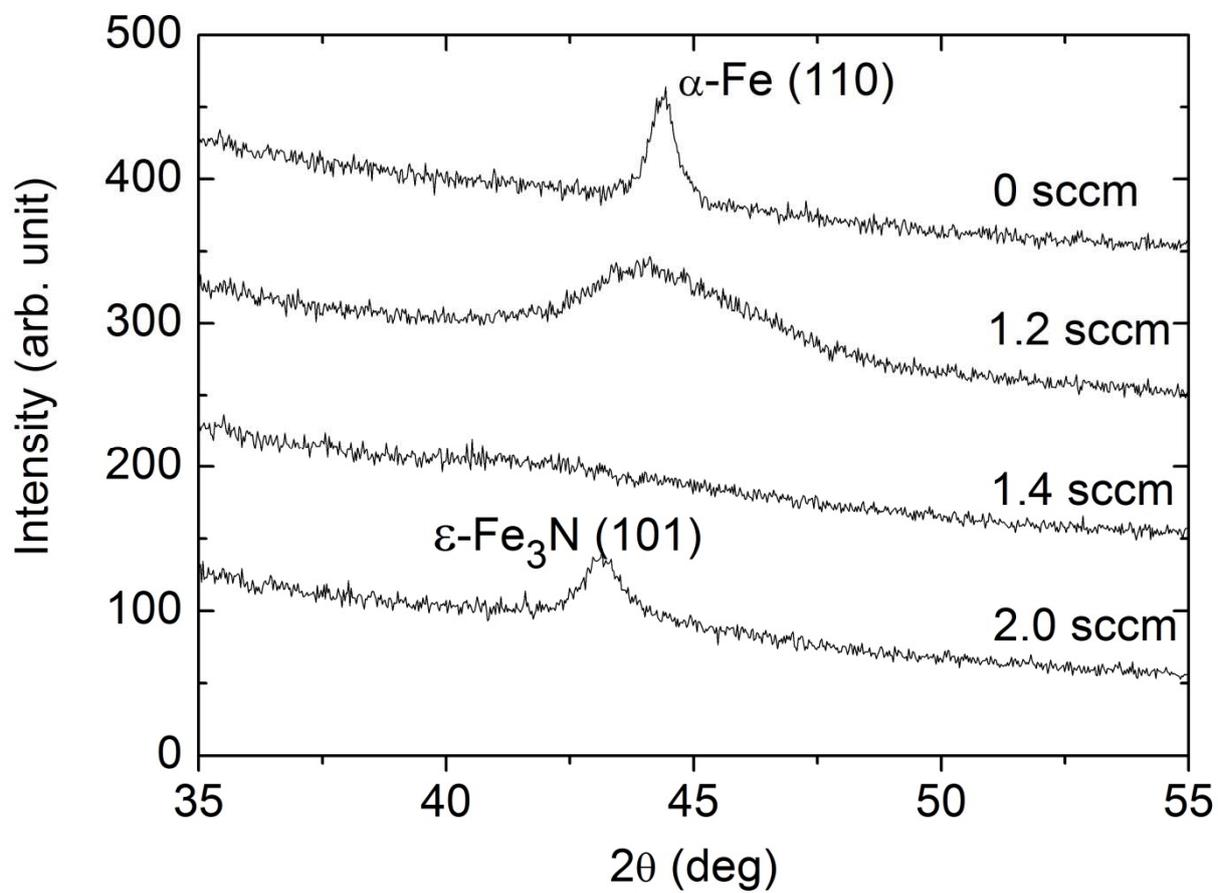

Fig. 1

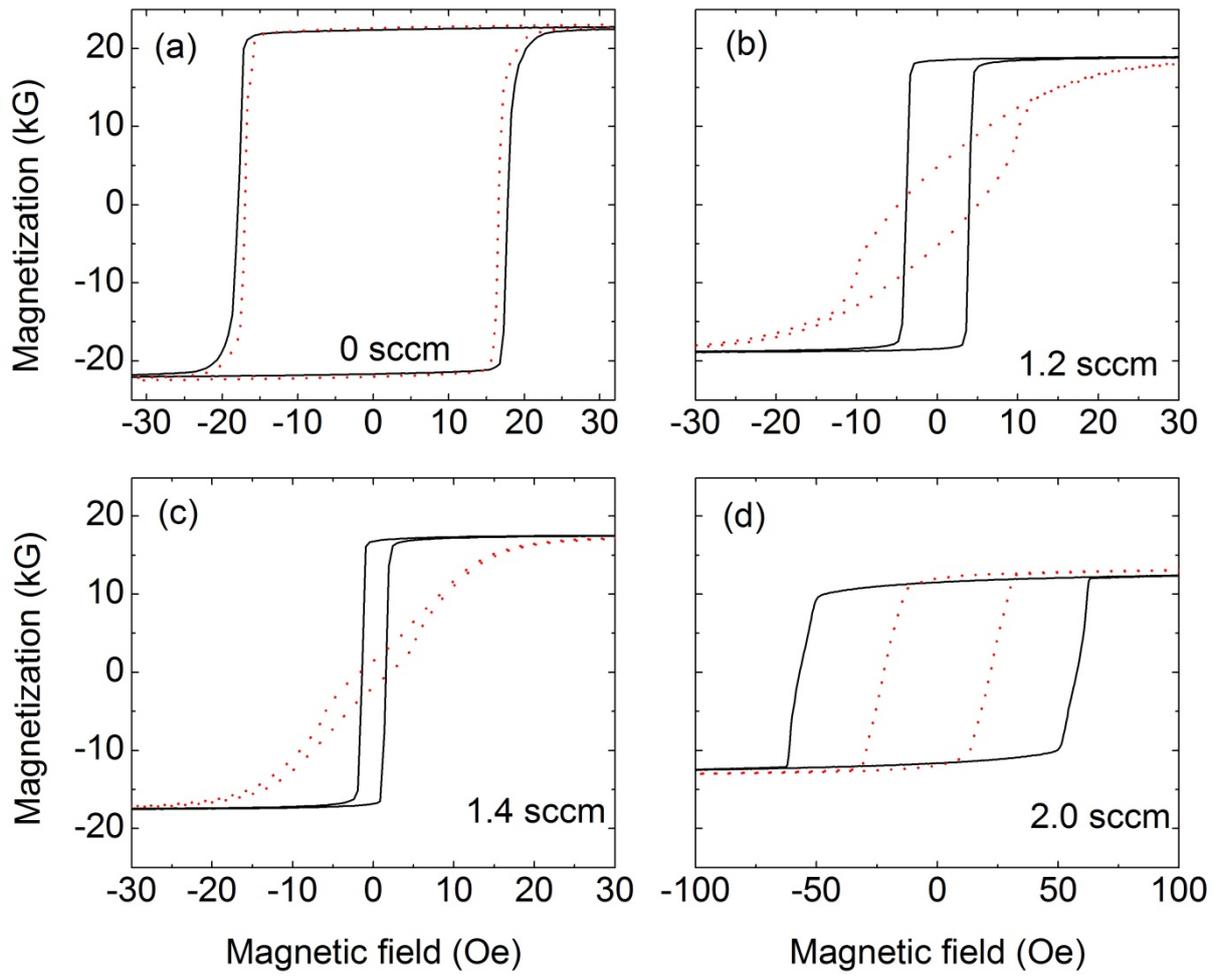

Fig. 2

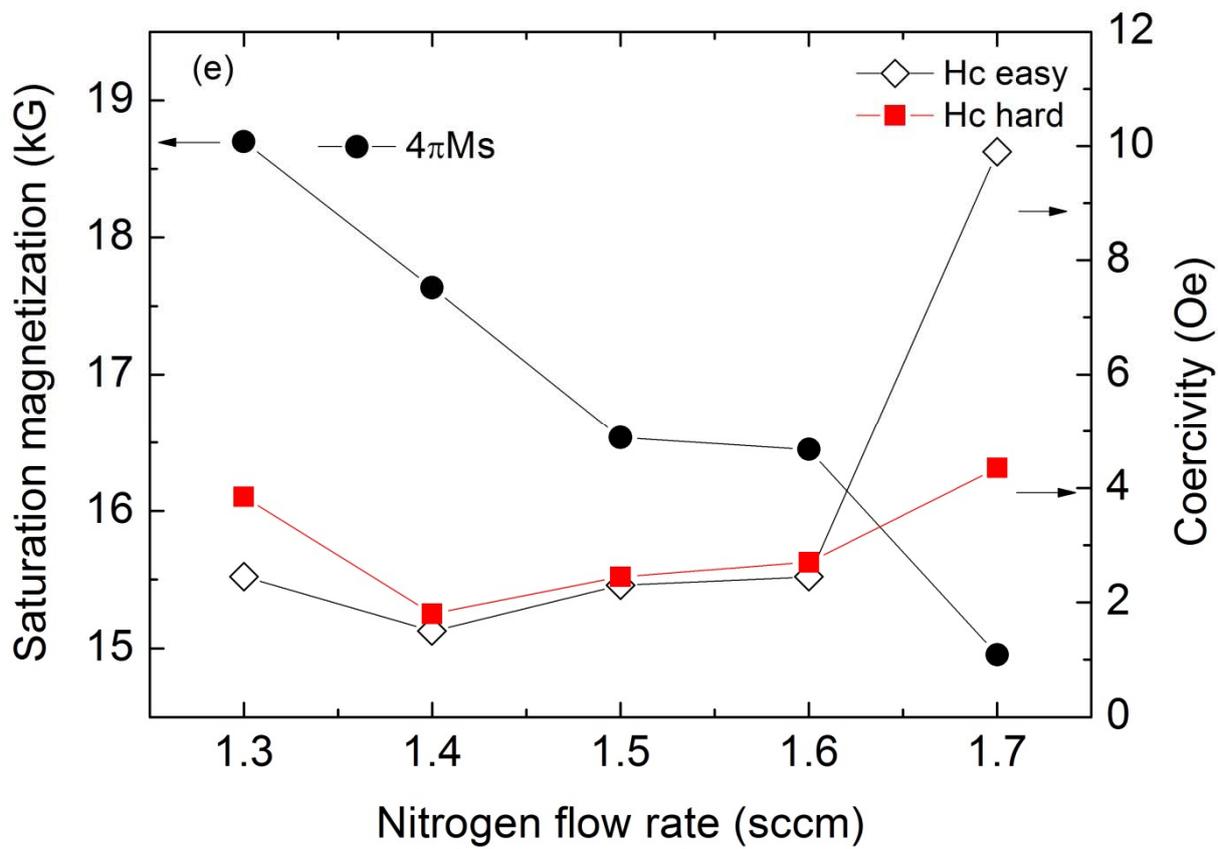

Fig. 3

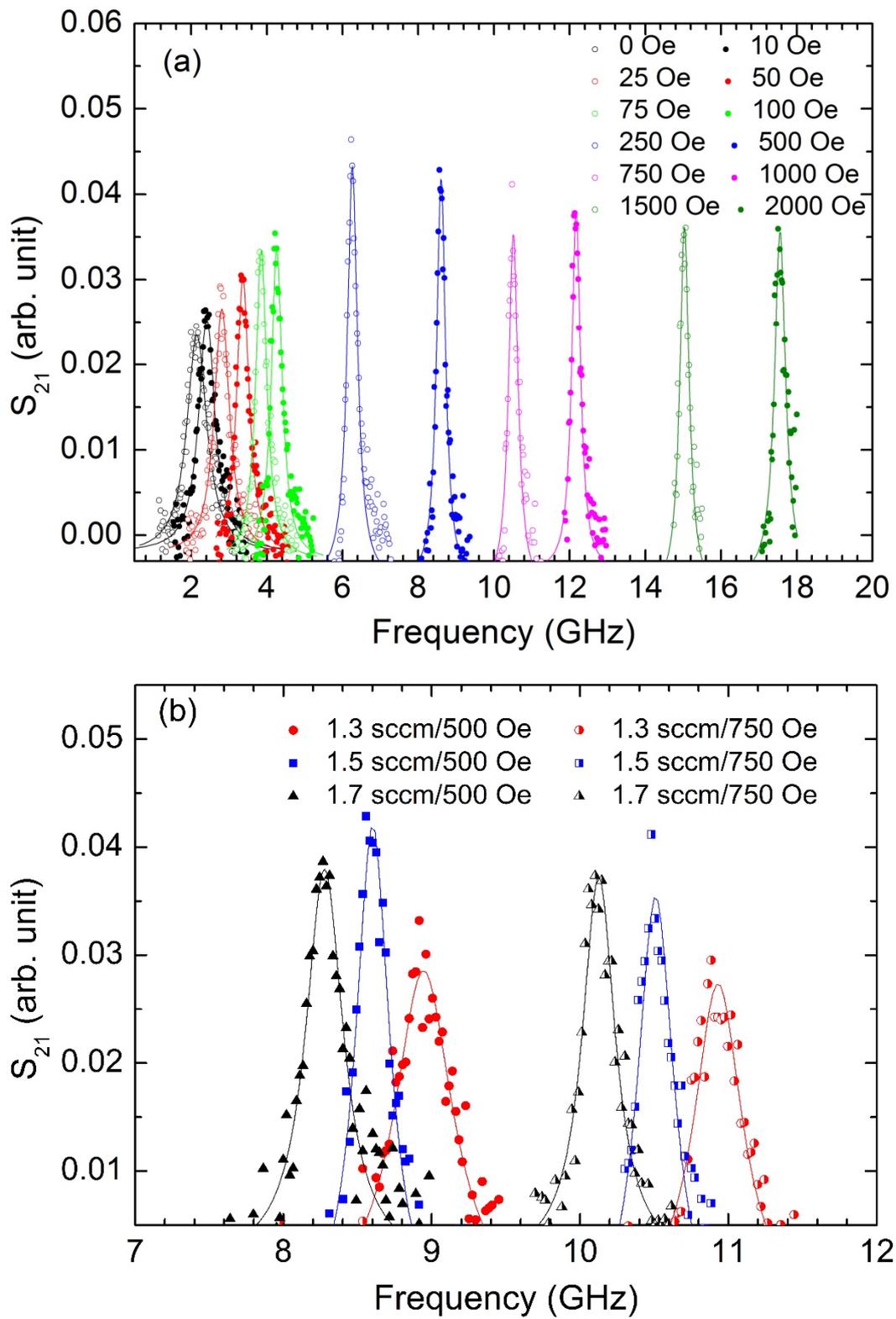

Fig. 4

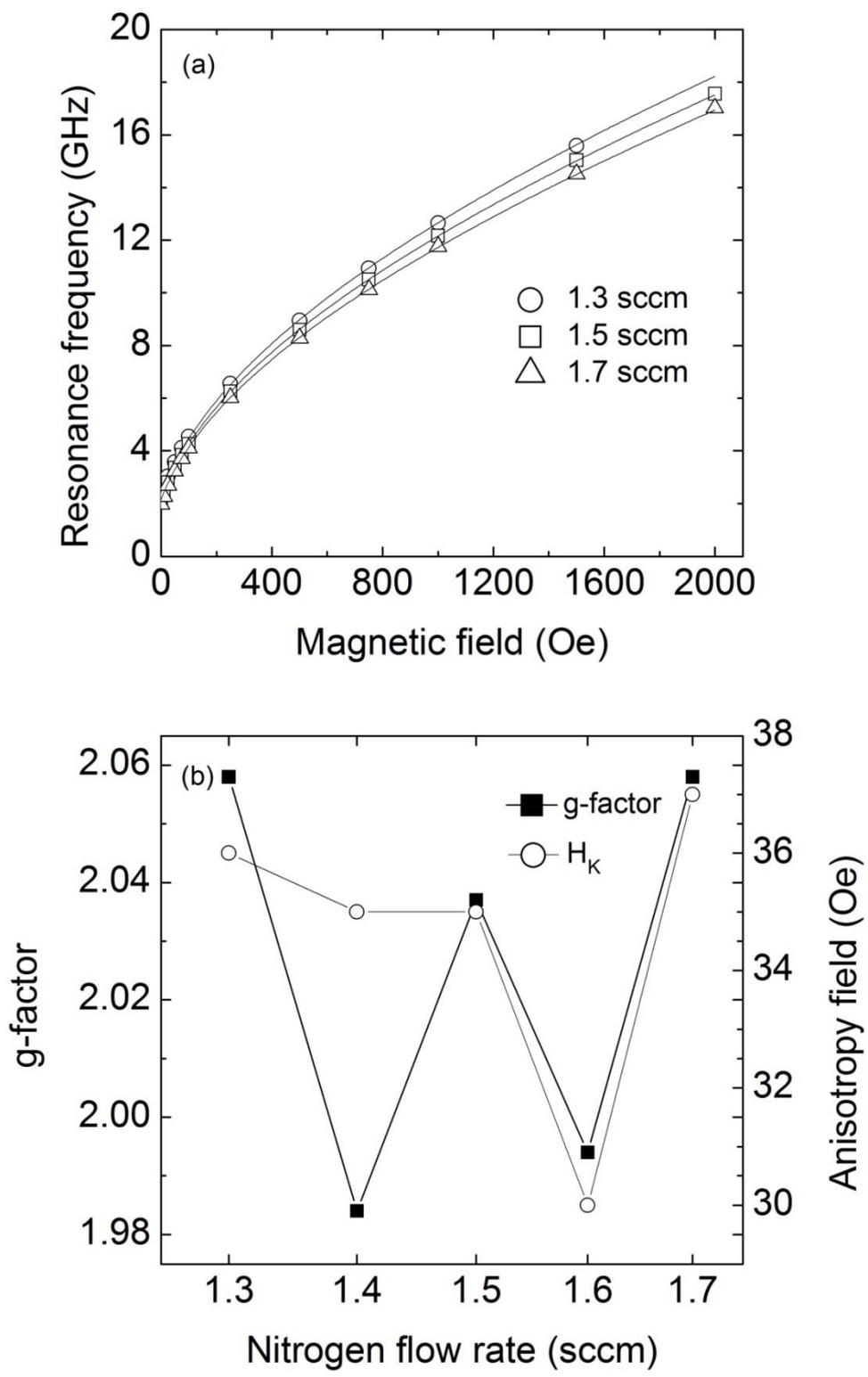

Fig. 5

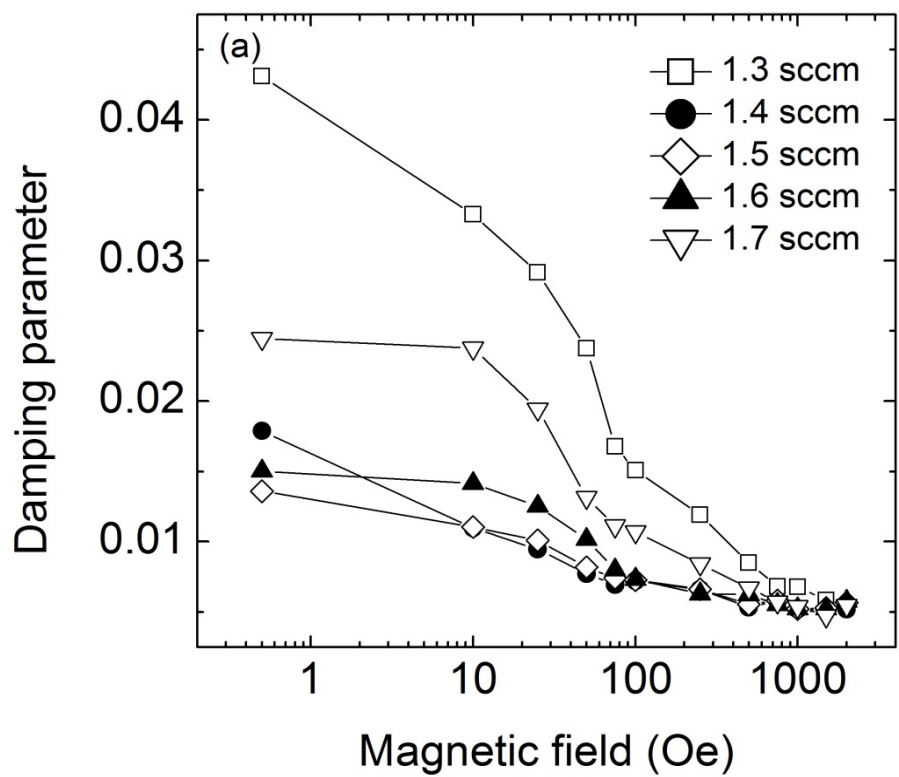

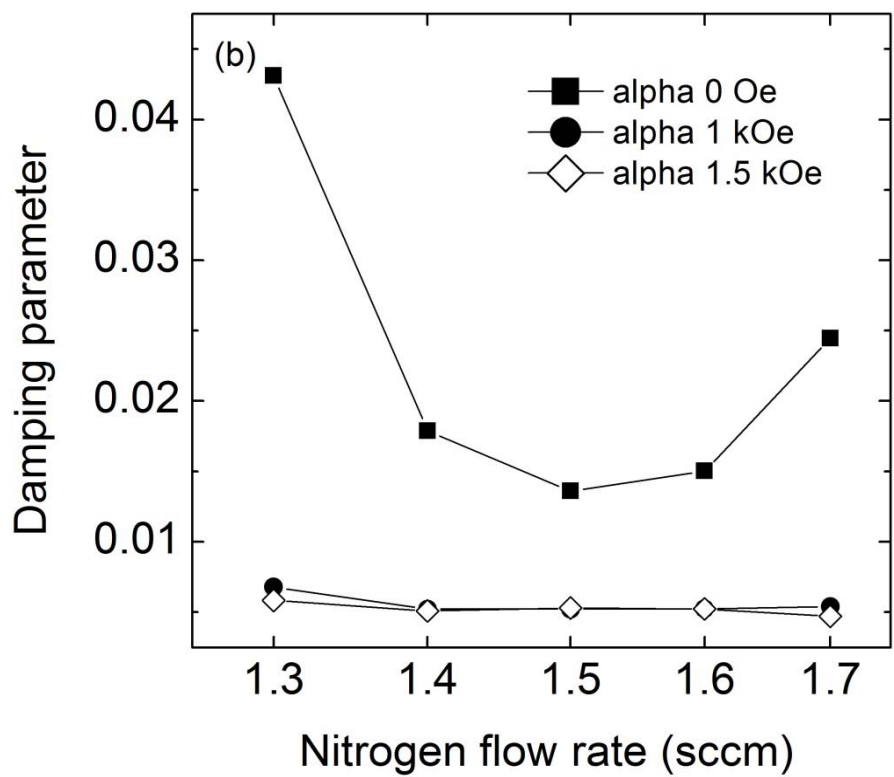

Fig. 6